\documentclass[11pt,bezier]{article}
\usepackage{amsfonts}

\title{The Improvement of the Bound on Hash Family}
\author{ Xianmin Feng,\hspace{3mm} Jiansheng
Yang\footnote{Corresponding author: Jiansheng Yang (1963~-~), Man,
Associate Professor, Doctor, Major in  Graph Theory, Coding Theory. E-mail: yjsyjs@staff.shu.edu.cn }\\ \\
School of Science, Shanghai University \\
 Shanghai 200444, China }
\date{}
\begin{document}
\maketitle

\begin{abstract}
In this paper, we study the bound on three kinds of hash family
using the Singleton bound. To $\varepsilon-U(N; n, m)$ hash family,
in the caes of $n>m^2>1$ and $1\geq\varepsilon\geq \varepsilon_1(n,
m)$, we get that the new bound is better. To
$\varepsilon-\bigtriangleup U(N; n, m)$ hash family , in the case of
$n>m>1$ and $1\geq\varepsilon\geq\varepsilon_3(n,m)$, the new bound
is better. To $\varepsilon-SU(N; n, m)$ hash family, in the case of
$n>2^m>2$ and $1\geq\varepsilon\geq \varepsilon_4(n, m)$, we get
that the new bound is better.

\medskip
{\em Keywords:} hash family;~the Singleton bound;~the Plotkin
bound;~MDS code
\end{abstract}

\section{Introduction}

\indent

The concept known as ``universal hashing" was invented by carter and
wegman [1] in 1979. In [2, p. 18], Avi Wigderson characterizes
universal hashing as being a tool which "should belong to the
fundamental bag of tricks every computer scientist". The hash
function has owned broad use in information authentication field
such as digital signature, and has had close relation to
authentication codes [10]. In 1980, D. V. Sarwate [5] introduced the
Plotkin bound to an $\varepsilon-U~(N; n, m)$ hash family, and got
$\varepsilon\geq \frac{n-m}{m(n-1)}$. In 1994, D. R. Stinson [3] got
$N\geq \frac{n(m-1)}{n(\varepsilon m-1)+m^2(1-\varepsilon)}$ when he
studied the $\varepsilon-U(N; n, m)$ hash family. In 1995, he [8]
got $N\geq\frac{n(m-1)}{m-n+m\varepsilon(n-1)}$ when studied the
$\varepsilon-\bigtriangleup U(N; n, m)$ hash family, and got $N\geq
1+\frac{n(m-1)^2}{m\varepsilon(n-1)+m-n}$ when studied the
$\varepsilon-SU(N;n,m)$ hash family.

In the following, We denote
$\frac{(n-m^2)(\log_mn-1))}{(mn-m^2)\log_mn+m^2+n-2mn}$,~$\frac{n-m}{m(n-1)}$,~\\
$\frac{n(m-1)(m-2)+(\log_2n+m-1)(n-m)}{m(n-1)(\log_2n+m-1)-2n(m-1)}$,
respectively, by $\varepsilon_1(n,m)$, $\varepsilon_2(n,m)$,
$\varepsilon_3(n,m)$ and denote the smaller solution of the
equation:
$2(n-1)x^2-[m(n-1)(\log_2n-3)+6n-2m-4]x+(m-2)(nm-2n+1)+(n-m)\log_2n=0$
by $\varepsilon_4(n, m)$.

In this paper, we introduce the Singleton bound to $\varepsilon-U(N;
n, m)$ hash family, and get $N\geq \frac{\log_mn-1}{\varepsilon}$.
Through comparing the two bounds, we get that the new bound is
better when $1\geq\varepsilon\geq \varepsilon_1(n, m)$, and the old
bound is better when $\varepsilon_1(n,
m)>\varepsilon\geq\varepsilon_2 (n, m)$. Meanwhile, we introduce the
Singleton bound to $\varepsilon-\bigtriangleup U(N; n, m)$ hash
family, and get $N\geq \frac{\log_2n+m-1}{m-2+2\varepsilon}. $
Through comparing, we get that the new bound is better when
$1\geq\varepsilon\geq \varepsilon_3(n, m)$, and the old bound is
better when $\varepsilon_3(n, m)>\varepsilon\geq\frac{1}{m}$. We
also introduce the Singleton bound to $\varepsilon-SU(N; n, m)$ hash
family, and get $N\geq \frac{m\log_2n}{m-2(1-\varepsilon)}$. Through
comparing, we get that the new bound is better when
$1\geq\varepsilon\geq \varepsilon_4(n, m)$, and the old bound is
better when $\varepsilon_4(n, m)>\varepsilon\geq\frac{1}{m}$.

\section{Hash Family and Codes}

\indent

{\bf Definition 2.1[7]}~~Let A, B are finit sets, suppose $|A|\geq
|B|$, the function $h:~A\rightarrow B$ is called hash function.

\medskip

{\bf Definition 2.2[3]}~~Let $\hbar$ is the set of hash function
$h:~A\rightarrow B$, if $|A|=n$, $|B|=m$, $|\hbar|=N$, then it is
called $(N; n, m)$ hash family.

\medskip

{\bf Definition 2.3[8]}~~An $(N; n, m)$ hash family is
$\varepsilon-universal$ provided that for any two distinct elements
$a_{1},a_2\in A$, there exist at most $\varepsilon N$ functions
$h\in \zeta$ such that $h(a_1)=h(a_2)$. we will use the notation
$\varepsilon-U$ as an abbreviation for $\varepsilon-universal$.

If the $\varepsilon$ of an $\varepsilon-U(N; n, m)$ hash family is
$\frac{1}{m}$, it is known as $universal$ hashing[1].

\medskip

 Generally, to an $\varepsilon-U(N; n, m)$ hash family,
$\varepsilon N$ is the smaller the better.

\medskip

{\bf Definition 2.4[8]}~~Suppose that functions in an $(N; n, m)$
hash family, $\hbar$, have range $B=G$, where $G$ is an additive
abelian group (of order $m$). $\hbar$ is called
$\varepsilon-\bigtriangleup universal$ provided that for any two
distinct elements $a_{1},a_2\in A$ and for any element $b\in G$,
there exist at most $\varepsilon N$ functions $h\in \hbar$ such
that$h(a_1)-h(a_2)=b$. We will use the notation
$\varepsilon-\bigtriangleup U$ as an abbreviation for
$\varepsilon-\bigtriangleup universal$.

\medskip

{\bf Definition 2.5[8]}~~An $(N; n, m)$ hash family is
$\varepsilon-strongly~universal$ provided that the following two
conditions are satisfied:

 1. for any element $a\in A$ and
amy element $b\in B$, there exist exactly $N/m$ functions $h\in
\hbar$ such that $h(a)=b.$

2. for any two distinct elements $a_1, a_2\in A$ and for any two
(not necessayily distinct) elements $b_1,b_2\in B$, there exist at
most $\varepsilon N/m$ functions $h\in \hbar$ such that
$h(a_i)=b_i,~i=1,2.$

\medskip

We will use the notation $\varepsilon-SU$ as an abbreviation for
$\varepsilon-strongly~universal.$

\medskip

{\bf Theorem 2.6[6]}~~If there exists an $(N,K,D,q)$ code, then
there exists a $(1-\frac{D}{N})-U(N; K, q)$ hash family. Conversely,
if there exists an $\varepsilon-U(N;n,m)$ hash family, then there
exists an $(N,n,(1-\varepsilon)N,m)$ code.

\medskip

 {\bf Theorem 2.7[8]}~~If there exists an $[N,k,D,q]$ code
$C$ with the property that $e=(1,\cdots,1)\in C$, then there exists
a $(1-\frac{D}{N})-\bigtriangleup U(N; q^{k-1}, q)$ hash family
defined over $F_q$.

\medskip

 {\bf Theorem 2.8[5]}~~If there exists
an $\varepsilon-U(N; n, m)$ hash family, then
$$\varepsilon\geq \frac{n-m}{m(n-1)}.$$

\medskip

{\bf Theorem 2.9[8]}~~If there exists an $\varepsilon-\bigtriangleup
U(N; n, m)$ hash family, then $$\varepsilon\geq\frac{1}{m}.$$

\medskip

 {\bf Theorem 2.10[8]}~~If there exists an
$\varepsilon-SU(N; n, m)$ hash family, then
$$\varepsilon\geq\frac{1}{m}.$$

\medskip

 {\bf Theorem 2.11[3]}~~If there exists an
$\varepsilon-U(N; n, m)$ hash family, then \begin{equation} N\geq
\frac{n(m-1)}{n(\varepsilon m-1)+m^2(1-\varepsilon)}.\end{equation}

\medskip

 {\bf Theorem 2.12[8]}~~If there exists an
$\varepsilon-\bigtriangleup U(N; n, m)$ hash family, then
\begin{equation}N\geq\frac{n(m-1)}{m-n+m\varepsilon(n-1)}.\end{equation}

\medskip

 {\bf Theorem 2.13[8]}~~If there exists an
$\varepsilon-SU(N; n, m)$ hash family, then \begin{equation}N\geq
1+\frac{n(m-1)^2}{m\varepsilon(n-1)+m-n}.\end{equation}

\medskip

The following discussion demands $m>1$.

\section{An New Bound for $\varepsilon-U$ Hash Family}

\indent

{\bf Theorem 3.1[4]}~~For $q, n, d \in N$, $q\geq 2$, we have $$A(n,
d)\leq q^{n-d+1}.$$

\medskip

So, we can get $K\leq q^{N-D+1}$ in an $(N, K, D, q)$ $(q\geq2)$
code. This is called the Singleton bound.

\medskip

{\bf Theorem 3.2}~~If there exists an $\varepsilon-U(N;n,m)$ hash
family, then
\begin{equation}N\geq \frac{\log_mn-1}{\varepsilon}.\end{equation}

\medskip

 \textbf{Proof:}~~From Theorem 2.6, since there exists an
$\varepsilon-U(N;n,m)$ hash family, then there exists an
$(N,n,(1-\varepsilon)N,m)$ code. Using the Singleton bound, we
get$$n\leq m^{N-(1-\varepsilon)N+1}.$$

So, $$\log_mn\leq \varepsilon N+1.$$

 Thus,$$N\geq\frac{\log_mn-1}{\varepsilon}.$$\hfill$\Box$

\medskip

{\bf Lemma 3.3}~~If $n>m^2$,
then$$1>\frac{(n-m^2)(\log_mn-1))}{(mn-m^2)\log_mn+m^2+n-2mn}>
\frac{n-m}{m(n-1)}.$$

\medskip

 \textbf{Proof:}~~ Let $\log_mn=1+\alpha$, since
$n>m^2>1$, then $\alpha>1$. Thus
$$\begin{array}{l}
~~~\frac{(n-m^2)(\log_mn-1))}{(mn-m^2)\log_mn+m^2+n-2mn}=\frac{(n-m^2)\alpha}{m(1+\alpha)(n-m)+m^2+n-2mn}\\=
\frac{(n-m^2)\alpha}{(\alpha-1)mn+n-m^2\alpha}=\frac{(n-m^2)\alpha}{(\alpha-1)(m-1)n+(n-m^2)\alpha}<1.
\end{array}$$

\medskip

Conside function: $f(x)=m^x-mx~(m>1)$,~$f^{'}(x)= m^{x}\ln m-m$.
Since $m>1$, then $f^{'}(x)>0$ when $x>1$, this means $f(x)$ is a
strictly monotony increasing function. Moveover, $f(x)=0$ when
$x=1$. That is to say, $m^\alpha>m\alpha$ when $n>m^2>1$.

 So, $n(m-1)m(-m\alpha+\alpha+m^\alpha-1)>0$ is true.

 Since $\log_mn=\alpha+1$ $(\alpha>1)$, then
$n=m^{\alpha+1}$, substitute it to the above inequality, we get
$$n(m-1)(-m^2\alpha+m\alpha+n-m)>0.$$

 Now, by this, we have
$$(n\alpha-m^2\alpha)(mn-m)>(mn\alpha-m^2\alpha-mn+n)(n-m).$$

 Since \vskip 5pt$mn(\alpha-1)-m^2\alpha+n=
mn(\alpha-1)-m^2(\alpha-1)+n-m^2= m(\alpha-1)(n-m)+n-m^2>0,$

 so,
$$\frac{(n-m^2)\alpha}{mn(\alpha-1)-m^2\alpha+n}>\frac{n-m}{m(n-1)}.$$

 Thus,
$$\frac{(n-m^2)(\log_mn-1))}{(mn-m^2)\log_mn+m^2+n-2mn}>\frac{n-m}{m(n-1)}.$$\hfill$\Box$

\medskip

 {\bf Theorem 3.4}~~If there exists an
$\varepsilon-U(N;n,m)$ hash family, and $n>m^2$, then the bound (4)
is better than (1) when $1\geq\varepsilon\geq\varepsilon_1(n,m)$;
the bound (1) is better than (4) when
 $\varepsilon_1(n,m)>\varepsilon\geq\varepsilon_2(n,m).$

\medskip

\textbf{Proof:}~~The bound (4) better than (1) means
$$\frac{\log_mn-1}{\varepsilon}\geq \frac{n(m-1)}{n(\varepsilon
m-1)+m^2(1-\varepsilon)}.$$

 From the above inequality, we have $$\varepsilon\geq
\frac{(n-m^2)(\log_mn-1))}{(mn-m^2)
\log_mn+m^2+n-2mn}=\varepsilon_1(n, m).$$

 Using the same way, we can get: the bound (1) is better
when $\varepsilon_1(n,m)>
\varepsilon\geq\varepsilon_2(n,m)$.\hfill$\Box$

\medskip

 {\bf Note:}~~In both bounds, $\varepsilon N$ must be an
integer.

\medskip

 {\bf Theorem 3.5}~~There exists an $\varepsilon-U(N;n,m)$
hash family and $N=\frac{\log_mn-1}{\varepsilon}$ if and only if
there exists an MDS code $ (N, n, (1-\varepsilon)N, m)$. \vskip 5pt
It is clear. Now, we have:

\medskip

 {\bf Theorem 3.6}~~Suppose $q$ is a power of prime with
$(1<k<n\leq q+1)$. Then there is a $\frac{k-1}{n}-U(n;q^k,q)$ hash
family.

\medskip

 Another nice application also uses MDS code. From [4], we
know there exists an $[n, n-1, 2, q]~(q\geq2)$ MDS code. Let
$n=q^{i+1}$, there exists its subcode $(q^{i+1}, (2q-1)\times
q^{q^{i+1}-3}, 2, q)$. Applying Theorem 2.6, the following is
obtained.

\medskip

 {\bf Theorem 3.7}~~There exists a
$(1-\frac{2}{q^{i+1}})-U(q^{i+1}; (2q-1)\times q^{q^{i+1}-3}, q)~
(q\geq2,~i\geq1)$ hash family.

\medskip

 From this theorem, $m=q,~n=(2q-1)\times
q^{q^{i+1}-3},~\varepsilon=1-\frac{2}{q^{i+1}}$, we have
$\varepsilon>\varepsilon_1(n,m)$, then the bound (4) is better. So,
$N\geq\lceil\frac{\log_mn-1}{\varepsilon}\rceil=\lceil[q^{{i+1}}-3-1+\log_q(2q-1)]\times
\frac{q^{i+1}}{q^{i+1}-2})\rceil=q^{i+1}$. Thus,
$(1-\frac{2}{q^{i+1}})-U(q^{i+1}; (2q-1)\times q^{q^{i+1}-3}, q)$
hash family has the smallest $N$.

\section{An New Bound for $\varepsilon-\bigtriangleup U$ Hash Family}

\indent

 {\bf Lemma 4.1}~~If $n>m$, then
$$1>\frac{n(m-1)(m-2)+(\log_2n+m-1)(n-m)}{m(n-1)(\log_2n+m-1)-2n(m-1)}>
\frac{1}{m}.$$

\medskip

 \textbf{Proof:}~~Since $n>m>1$, then
$(m-1)n(1-\log_2n)<0$. We have
$$n(m-1)(m-2)+(\log_2n+m-1)(n-m)<m(n-1)(\log_2n+m-1)-2n(m-1).$$

 Since
$$m(n-1)(\log_2n+m-1)-2n(m-1)=m(n-1)(\log_2n+m-3)+2n-2m>0,$$

 thus,
$$1>\frac{n(m-1)(m-2)+(\log_2n+m-1)(n-m)}{m(n-1)(\log_2n+m-1)-2n(m-1)}.$$

 Since $$m(m-1)(n-\log_2n)+(m-1)^2(mn-m-2n)>0,$$

 then

$mn(m-1)(m-2)+m(\log_2n+m-1)(n-m)-[m(n-1)(\log_2n+m-1)-2n(m-1)]>0.$

 Thus,
$$\frac{n(m-1)(m-2)+(\log_2n+m-1)(n-m)}{m(n-1)(\log_2n+m-1)-2n(m-1)}>\frac{1}{m}.$$\hfill$\Box$

\medskip

 From the proof of Theorem 2.12 in [8], we have that if
there exists an $\varepsilon-\bigtriangleup U(N; n, m)$ hash family,
then there exists a constant-weight $((N-1)m, n, 2N(1-\varepsilon),
2)$ code. Using the Singleton bound, we have
$$n\leq2^{(N-1)m-2(1-\varepsilon)N+1}.$$ So, the bound (4) is
changed to \begin{equation} N\geq
\frac{\log_2n+m-1}{m-2+2\varepsilon}.\end{equation}

\medskip

 {\bf Theorem 4.2}~~If there exists an
$\varepsilon-\bigtriangleup U(N; n, m)$ hash family and $n>m$, then
the bound (5) is better than (2) when
$1\geq\varepsilon\geq\varepsilon_3(n,m)$; the bound (2) is better
than (5) when $\varepsilon_3(n,m)>\varepsilon\geq\frac{1}{m}.$

\medskip

 \textbf{Proof:}~~The bound (5) better than (2) means
$$\frac{\log_2n+m-1}{m-2+2\varepsilon}\geq
\frac{n(m-1)}{m-n+m\varepsilon(n-1)}.$$

 From the above inequality, we have
$$\varepsilon\geq \frac{n(m-1)(m-2)-(\log_2n+m-1)(m-n)}{m(n-1)(\log_2n+m-1)-2n(m-1)}=\varepsilon_3(n, m).$$

 Using the same way, we can get: the bound (2) is better
when $\varepsilon_3(n,m)> \varepsilon\geq\frac{1}{m}$.\hfill$\Box$

\medskip

 {\bf Note:}~~In both bounds, $\varepsilon N$ must be an
integer.

\medskip

 {\bf Example:}~~Let
$\varepsilon=1-\frac{2}{(q-1)^{i+1}}~(q>2,~i>1),~n=q^{{(q-1)}^{i+1}-1},~m=q$,
then through computing, we have $1>\varepsilon>\varepsilon_3(n,m)$,
so the bound (5) is better. Then $$N\geq
\frac{\log_2n+m-1}{m-2+2\varepsilon}=\frac{(q-1)^{i+1}\log_2q+q-\log_2q-1}{q+\frac{4}{(q-1)^{i+1}}}.$$

 Since $q>2,~i>1$, then we have
$$(q-1)^i<\frac{(q-1)^{i+1}
\log_2q+q-\log_2q-1}{q+\frac{4}{(q-1)^{i+1}}}<(q-1)^{i+1}.$$

 Since $\varepsilon N$ is an integer, then $N\geq
(q-1)^{i+1}.$

 From [4], we know there exists an $[n, n-1, 2, q]$ MDS
code $C$. Let $n=(q-1)^{i+1}$, we may assume that $e=(1,\cdots,1)\in
C$, then from Theorem 2.7, we have the following.

\medskip

 {\bf Theorem 4.3}~~ There exists a
$(1-\frac{2}{(q-1)^{i+1}})-\bigtriangleup U((q-1)^{i+1};
q^{(q-1)^{i+1}-1}, q)$\\$(q>2,~i>1)$ hash family.

\section{An New Bound for $\varepsilon-SU$ Hash Family}

\indent

 For $n,~m>0$, we denote $2(n-1)$ by $a$, denote
$m(n-1)(\log_2n-3)+6n-2m-4$ by $b$ and denote
$(m-2)(nm-2n+1)+(n-m)\log_2n$ by $c$. Then, $$\varepsilon_4(n,
m)=\frac{b-\sqrt{b^2-4ac}}{2a}.$$

\medskip

 {\bf Lemma 5.1}~~If $n>2^m$, then $$1>\varepsilon_4(n,
m)>\frac{1}{m}.$$

\medskip

 \textbf{Proof:}~~Since $n>2^m>1$, then
$n(m-1)(m-\log_2n)<0$. So,

$2(n-1)-[m(n-1)(\log_2n-3)+6n-2m-4]+(m-2)(nm-2n+1)+(n-m)\log_2n<0.$

 This is to say $a+b+c<0$. Thus, $$\varepsilon_4(n,
m)=\frac{-b-\sqrt{b^2-4ac}}{2a}<1.$$

 Since

$2(n-1)-m[m(n-1)(\log_2n-3)+6n-2m-4]+m^2[(m-2)(nm-2n+1)+(n-m)\log_2n]
=2(n-1)+m^2\log_2n+m(m^2-3m+4)+3mn(m-2)+m^3(n-\log_2n)+m^2n(m-1)(m-4)>0.
$

 (It is obvious when $m\geq4$. So, we only need to check
on the cases $m=2$ and $m=3$ to get the result.)

 This is to say, $$a+bm+cm^2>0.$$

 Thus,
$$\varepsilon_4(n, m)>\frac{1}{m}.$$\hfill$\Box$

\medskip

 From the proof of Theorem 2.13 in [8], we have that if
there exists an $\varepsilon-SU(N; n, m)$ hash family, then there
exists a constant-weight $(N-1, n, 2(1-\varepsilon)N/m, 2)$ code.
Using the Singleton bound, we have
$$n\leq2^{N-1-2N(1-\varepsilon)/m+1}.$$ So, the bound (4) is changed
to \begin{equation} N\geq
\frac{m\log_2n}{m-2(1-\varepsilon)}.\end{equation}

\medskip

{\bf Theorem 5.2}~~If there exists an $\varepsilon-SU(N; n, m)$ hash
family and $n>2^m$, then the bound (6) is better than (3) when
$1\geq\varepsilon\geq\varepsilon_4(n,m)$; the bound (3) is better
than (6) when $\varepsilon_4(n,m)>\varepsilon\geq\frac{1}{m}.$

\medskip

 \textbf{Proof:}~~The bound (6) better than (3) means
$$\frac{m\log_2n}{m-2(1-\varepsilon)}\geq 1+\frac{n(m-1)^2}{m\varepsilon(n-1)+m-n}.$$

 From the above inequality, we have $$1\geq\varepsilon\geq
\varepsilon_4(n,m).$$

 Using the same way, we can get: the bound (3) is better
when $\varepsilon_4(n,m)> \varepsilon\geq\frac{1}{m}$.\hfill$\Box$

\medskip

 {\bf Note:}~~In both bounds, $\frac{\varepsilon N}{m}$
must be an integer.

\end{document}